\documentclass[preprint,nocopyrightspace]{sigplanconf}

\usepackage{amsmath}

\usepackage{blindtext, graphicx}
\usepackage{graphicx}
\usepackage{caption}
\usepackage{subcaption}
\usepackage{booktabs}
\usepackage{tabularx}
\usepackage{comment}
\usepackage{subfig}
\usepackage[british]{babel}
\usepackage{comment}
\usepackage{floatrow}
\newfloatcommand{capbtabbox}{table}[][\FBwidth]
\usepackage{multirow}

\begin{document}

\setlength{\pdfpageheight}{\paperheight}
\setlength{\pdfpagewidth}{\paperwidth}



\preprintfooter{PROHA'16, March 12, 2016, Barcelona, Spain}

\title{ If-Conversion Optimization using Neuro Evolution of Augmenting Topologies }

\authorinfo{Reem Elkhouly}
           {Computer Science and Eng. Dept.,\\E-JUST, Alex., Egypt}
           {reem.elkhouly@ejust.edu.eg}
\authorinfo{Keiji Kimura}
           {Dept. of Computer Science and Eng. \\Waseda University, Tokyo, Japan}
           {kimura@apal.cs.waseda.ac.jp}           
           
\authorinfo{Ahmed El-Mahdy}
           {Computer Science and Eng. Dept.,\\E-JUST, Alex., Egypt}
           {ahmed.elmahdy@ejust.edu.eg}

\maketitle

\begin{abstract}
Control-flow dependence is an intrinsic limiting factor for program acceleration. With the availability of instruction-level parallel architectures, if-conversion optimization has, therefore, become pivotal for extracting parallelism from serial programs. While many if-conversion optimization heuristics have been proposed in the literature, most of them consider rigid criteria regardless of the underlying hardware and input programs.
In this paper, 
we propose a novel if-conversion scheme that preforms an efficient if-conversion transformation using a machine learning technique (\emph{NEAT}). This method enables if-conversion customization overall branches within a program unlike the literature that considered individual branches. Our technique also provides flexibility required when compiling for heterogeneous systems.
%
The efficacy of our approach is shown by experiments and reported results which illustrate that the programs can be accelerated on the same architecture and without modifying the original code. Our technique applies for general purpose programming languages (e.g. C/C++) and is transparent for the programmer.
We implemented our technique in LLVM 3.6.1 compilation infrastructure and experimented on the kernels of SPEC-CPU2006 v1.1 benchmarks suite running on a multicore system of Intel(R) Xeon(R) 3.50GHz processors.
Our findings show a performance gain up to 8.6\%  over the standard optimized code (LLVM -O2 with if-conversion included), indicating the need for If-conversion compilation optimization that can adapt to the unique characteristics of every individual branch. 

\end{abstract}


\section{Introduction}

Fast performing programs represent a chief goal in the fields of computer architecture and compilers. A key performance driver is providing parallel execution at various levels of granularity. With the multicore shift, larger parallelism granularity is generally sought. In addition to that, single-core performance is still important, as it is a major scaling limiting factor. 
Accelerating single-core performance mainly relies on instruction-level parallelism (ILP)~\cite{wall1991limits}, where various independent instructions are executed in the same processing cycle simultaneously on the same CPU (i.e. on different ALUs). The degree of parallelism is inherently limited by the data-flow characteristics of the running program. However, control-flow significantly hinders exploiting the `true dependence' manifested by the data-flow, significantly reducing the achievable degree of parallelism~\cite{Allen:1983:CCD:567067.567085}. 

One major approach to alleviating control-flow dependence is branch prediction, where the target of the branch is predicted using one of the multiple conventional branch predictors. Specifically, a branch predictor guesses whether the conditional branch causes a transfer of control or not~\cite{smith1981study}. It relies on the branching outcome history of the branch and/or other spatially contiguous branches. The more complicated the predictor is, the better guessing it can make. That alleviates the speculative execution in modern pipelined processors~\cite{1998dynamic}.

However, not all branches can be easily predicted, particularly those with a random outcome (such as branches relying on element comparisons when sorting random inputs). Mispredictions come with high time penalty resulting from the wasted processing cycles in executing the wrongly predicted instructions, flushing them and fetching the correct ones~\cite{hennessy2011computer}. 
Another argument that makes the branches undesirable is that they limit the instruction fetch bandwidth~\cite{shen2013modern} which diminishes the instruction level parallelism (ILP), hence the number of instructions executed per cycle.

The if-conversion approach is used to convert control-depen\-dence into data-depen\-dence, aiming to defeat the above performance obstacles. In this model, instructions are guarded by `predicates', if supported by the hardware otherwise conditional moves are used, thereby eliminating control-flow~\cite{mahlke1995comparison}. This approach thus relies on `if-conversion' optimizations to convert conditional branches into predicated instructions, allowing for further potential parallelization subject to the inherent data-flow dependences. However, predication comes at the extra cost of executing `nullified' instructions. That can potentially degrade performance for large `if-then' bodies. Moreover, branches interact in terms of allowing for different execution schedules, for which finding the optimal schedule is a hard combinatorial search problem where its complexity grows exponentially with the number of branches per function in the program.


In this paper, we revisit the problem of deciding which bran\-ches to convert. In particular, we implement the machine learning method \textit{NEAT}~\cite{stanley:ec02} in the LLVM compiler to replace the heuristics used in if-conversion optimization. The advantage of using this algorithm is the ability to evaluate the if-conversion for all branches in the program taking into consideration mutual influence.
\textit{NEAT} fits for this problem because of its high performance in searching large spaces (exponential in number of branches in this problem).
Our system successfully achieved up to 8.6\% performance gain transparently on the same architecture and does not require altering the original code by any mean. Our technique identifies code features that characterize different branches in the program. These features are fed to the machine learning algorithm \textit{NEAT} during the if-conversion optimization to customize the transformation plan.

Our contributions can be summarized in three basic modules: 
\begin{itemize}
\item First, a module that analyzes the code to capture code features as a vector for each branch. 
\item Second, a module that applies the \textit{NEAT} algorithm to repeatedly customize the if-conversion decision for all the branches as a vector of ones and zeros which we call the \textbf{bitmask} and eventually provides the best performing optimized program. 
\item Third, a module that controls the if-conversion optimization in the LLVM according to the bitmask generated by the second module. 
\end{itemize}
We consider C/C++ kernels from the SPEC-CPU2006 v1.1 ben\-chmarks suite~\cite{spec} for the experiments. 
The results are comprehended in comparison to the LLVM's compilation behavior on the same kernels. 
%

The rest of this paper is organized as follows; 
Section~\ref{background} explains essential background. Section~\ref{system} provides detailed description for our system and contributions.
Section~\ref{expr} presents and discusses results. Section~\ref{related} provides related work. Finally, Section~\ref{conclusion} concludes the paper and discusses future work.
\section{Background}
\label{background}
In this section, we provide the related background that is necessary to illustrate our system. 
Section~\ref{if_conv} presents various if-\-conversion transformations. Section~\ref{LLVM} introduces the LLVM if-\-conversion technique and how we use it to build our analysis module. Section~\ref{neat} explains the machine learning algorithm we used for our system (\textit{NEAT}).

\subsection{If-Conversion Optimizations}
\label{if_conv}
Control-flow optimizations, such as branch optimizations, relax the control dependences in the program in order to facilitate instruction parallelization and speculative execution. Conditional branches add control-flow dependences to the program, as their outcome is not known until runtime.
Systematically converting control dependence to data dependence was initially introduced by Allen et al.~\cite{Allen:1983:CCD:567067.567085}.
The if-conversion process aims to remove branches from a given program~\cite{OCMA02}. Certainly, a branch cannot be deleted without being replaced by another control that maintains program proper functionality. That could be accomplished either by conditional moves (CMOVs) or guarded (predicated) execution. 

\textbf{\textit{Conditional Moves (CMOVs):}}
They are hardware supported instructions that copy a value from a source to a destination if and only if a specific condition is true. This condition is assessed, and then one of the special purpose registers (i.e. EFLAGS register in x86) is changed.
That enables the compiler to convert a piece of code that contains simple branches into branch free code. Several processor architectures, such as the SPARC architecture starting from SPARC-V9 and Pentium Pro~\cite{ACDI97} and more recent processors, support conditional moves.

\textbf{\textit{Guarded (Predicated) Execution:}}
It removes forward branches by adding guard expressions that control the execution of the instructions. The guard is a Boolean expression that is constructed by joining a set of conditions; each of these represents a conditional branch that leads to the current instruction. It is obvious that the guard expression may grow complex if the instruction is control-dependent on many conditional branches. The guard expression could be simplified, but the problem of simplifying a Boolean expression is NP-Complete. Some processor architectures such as Intel IA-64 and ARM 32-bit support predicated execution by a fully predicated instruction set using special registers.
To ensure the correctness of the program after branch removal, three constraints should apply~\cite{OCMA02}:
(1) The guard expression of an instruction is true if and only if the corresponding statement in the original program is executed.
(2) The order of execution of instructions that have true guards in the new program is the same as their order of execution in the original program.
(3) Any expression with side effects is evaluated exactly as many times in the new program as the original program.
If-conversion optimizations are of surpassing importance because conditional branch instructions are very expensive --in terms of runtime-- when mispredicted, besides, being obstacles in the way of \textit{ILP}. 
 
We propose a technique for tunning if-conversion optimization - using LLVM - valid for any architecture that supports predicated execution. our technique uses a machine learning algorithm that is trained to tune if-conversion to give the best performance for the whole compiled program. When compiles for heterogeneous systems, different compilations are done for different underlying hardware ensuring a suitable optimization tuning for each of them. While the effect of branches and branch predictors differ among the architectures, our technique which is adaptable will be of great value.
The following section studies the utility of these optimizations in the LLVM if-conversion module, then, how we used it in our technique. 




\subsection{LLVM If-conversion}
\label{LLVM} 
The LLVM~\cite{llvm} is a compiler infrastructure that conducts code analysis and transformation using reusable modules, called passes, that support static and dynamic compilation. It includes a  wide range of capabilities that appear transparent to programs. Moreover, it provides high-level information at compile-time, link-time, runtime and between runs. Besides, LLVM provides a large number of analysis and optimization tools that enable precisely generating the object code in a way that serves specific endeavors. In this infrastructure, a low-level Static Single Assignment (SSA) form is elaborated for code representation~\cite{lattner2004llvm}.
LLVM provides if-conversion optimization either 
for the architectures that support predicated instructions
or the out-of-order CPUs, where both the ``then'' and ``else'' bodies are executed, and the result is chosen by a~\textit{cmov} instruction. Therefore, branches that may be mis-predicted are omitted. 

The LLVM if-conversion transformation is included in the -O2 optimization level and can also be separately requested using command line options. For every function, it traverses the whole dominator tree of basic blocks post-orderly. All instructions within the basic block should be valid for speculative execution in order to allow conversion. For the nested branches, the conversion starts from the most inner tracing back up to the outer ones. To avoid critical edges, the if-conversion pass handles either triangle or diamond branches only. Another conversion constraint is that the considered block should end with a convertible to ``select'' instruction,\textit{PHI} node. %
All of these measures would assure the program's correctness.

After checking the eligibility of if-conversion for a basic block, a heuristics based cost model is applied to decide if the conversion is profitable. The profitability is measured in terms of multiple criteria: First, the critical path should not be extended by more than a certain limit, chosen to be half of the misprediction penalty. 
Second, if-converted trace length shortening that is limited by ILP resources should exceed that of the critical path in addition to the misprediction before conversion. 
Finally, the delay to ``select'' instruction by data dependence in both ``then'' and ``else'' traces should be considered. 
In our system, we replace the cost model and heuristics by a machine learning algorithm (\textit{NEAT}) that decides the if-conversion profitability according to the code features and a corresponding fitness value reflects the performance of the whole program.
\begin{figure*}[!t]
\centering
\includegraphics[width=\textwidth]{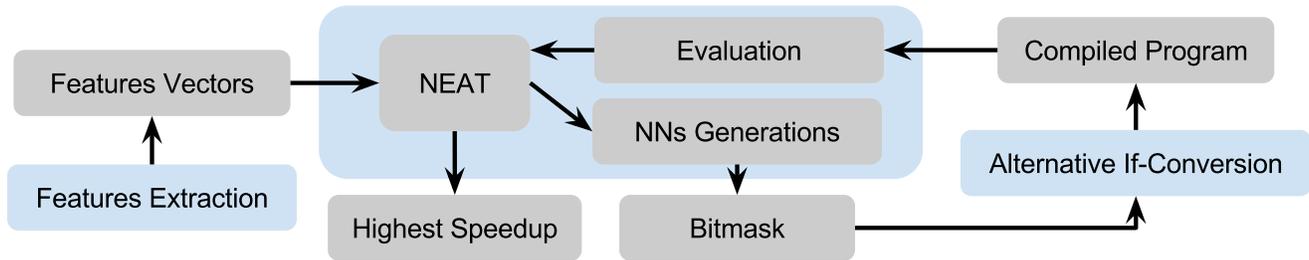}
\caption{System architecture that mainly consists of three modules.}
\label{fig:systemModules}
\end{figure*}
\subsection{NEAT}
\label{neat}
Neural networks are famous for their ability to model nonlinear problems of high complexity. NeuroEvolution of Augmenting Topologies (\textit{NEAT}) is a machine learning algorithm that artificially evolves neural networks (NNs) using genetic algorithms techniques. It does not only evolve the weights of connections, but also evolves the structure of the network. \textit{ NEAT} is proved to outperform fixed-topology techniques and showed promising performance in the reinforcement learning problems~\cite{stanley:ec02}. 

\textit{NEAT} starts from an initial generation of minimal uniform NNs where the input nodes are connected directly to the output nodes with zero hidden nodes.  Genetic encoding is used where genomes linearly represent NNs. Thus a genome is a list of connection genes, each connecting two node genes and specifies the corresponding weight. \textit{NEAT} allows a mutation that can change connection weights and network structure (adding/deleting nodes and/or connections) to generate a population of NNs. The outputs of these networks are evaluated through a fitness function, then the best of them are selected to be the parents for the next generation of networks~\cite{stanley1996efficient}. This process is repeated over successive generations of NNs, preserving dimensionality minimization through minimal structure incremental growth, until the desired fitness is found. In our system, we use \textit{NEAT} as a machine intelligence technique to search the space of $2^n$ possible combinations of converted and non-converted $n$ branches in the program in order to find the combination that achieves the highest performance.


\section{System Description}
\label{system}
\begin{table*}[!ht]
\centering
\begin{tabular}{@{}l l@{}}
\toprule
Feature&Description\\
\midrule
\textbf{\textit{Basic Block (BB)} size}&No. of instructions in the \textit{Basic Block (BB)} that contains the branch.\\
\textbf{True \textit{BB} Critical Path}& Critical Path length -data dependency- to the True \textit{BB}.\\
\textbf{False \textit{BB} Critical Path}& Critical Path length -data dependency- to the False \textit{BB}.\\
\textbf{Minimal Critical Path}&Minimum of the \textit{True} and \textit{False} Critical Paths.\\
\textbf{Unexploited \textit{ILP}}&Maximum \textit{ILP} achieved within the \textit{BB} after if-conversion.\\
\textbf{Branch Depth}&The distance of the earliest issue cycle as determined by data dependences \\
 &and latencies from the beginning of the trace.\\
\textbf{Loop Depth}&The nesting level of the loop that contains the \textit{BB}.\\
\textbf{\textit{Slack Sum}}&No. of \textit{Cycles} the instruction can be delayed without increasing \textit{Critical Path}.\\
\textbf{Maximum depth}&The \textit{slack} of the \textit{PHI} node in addition to its depth from the tail trace beginning.\\
\textbf{True \textit{BB} depth}&The depth of the \textit{PHI} node from the True \textit{BB} + \textit{PHI True Cycles}.\\
\textbf{False \textit{BB} depth}&The depth of the \textit{PHI} node from the False \textit{BB} + \textit{PHI False Cycles}.\\
\bottomrule
\end{tabular}
\caption{Code features used by \textit{NEAT} to customize if-conversion}
\label{tab:features}
\end{table*}
Our system customizes the if-conversion optimization replacing the fixed heuristics used in literature to estimate the profitability of the if-conversion by a machine learning strategy based on \textit{NEAT}. It works for any program written in any language which is supported by the LLVM compiler and does not require any changes or additions to the original code of the program.
We apply \textit{NEAT} to a set of descriptive features that precisely characterizes the code of \textit{Basic Blocks (BBs)} that contain different branches in the program. For this purpose, our system is implemented in a three major modules illustrated in Figure~\ref{fig:systemModules} which are implemented as plugins in the compiler infrastructure. 

\textbf{Module 1} is responsible for inspecting the code to find the branches that can be converted safely without affecting the correctness of the program, then extracts the set of features listed in Table~\ref{tab:features} for each one of them. These features are stored in vectors then forwarded to the next module. 

\textbf{Module 2} receives the features vectors and pushes them as an input for the \textit{NEAT} initial population of \textit{NNs}. Each \textit{NN} is responsible for evaluating the features of a single branch to decide whether it shall be if-converted. This decision is expressed as a single \textit{NN} output (1/0). The output of all \textit{NNs} - a stream of 1's/0's which we call a \textbf{bitmask} - is sent to the third module that controls the if-conversion of the corresponding branches on/off. The program is compiled with the customized if-conversion optimization, executed and evaluated against the fitness function which we designed as the speedup over the program when optimized with the original if-conversion of the LLVM. A group of the \textit{NNs} that gave the best performance are chosen to be the parents of the next generation of \textit{NNs}. A new generation of \textit{NNs} is generated from those parents using \textit{GA} which enhances the structure of \textit{NNs}. \textit{NEAT} continues to iterate over new generations to maximize the fitness (speedup) until it reaches a defined limit of number of generations. 

\textbf{Module 3} is an alternative for the original if-conversion \textit{pass} in LLVM. This module receives the \textbf{bitmask} for every program compilation where the number of bits equals the number of branches inspected. The if-conversion for each branch is turned on/off according to the value of the corresponding bit. This if-conversion is done within the whole compilation process, the output program is run and the speedup is calculated. The overall system eventually reports the highest speedup, the corresponding \textit{NN}, the bitmask and the best performing compiled version of the program. 

Our system is convenient for compiling for heterogeneous systems requires consideration to its special nature. Exploiting available resources as the hardware configuration changes is essential to make full utilization of them. The compiler should be aware of generating code that  targets different hardware. Our technique adjusts if-conversion optimization to suit targets. As it does not consider rigid hardware specifications as a metric for the optimization tuning. On the contrary, the tunning is done through learning from several runnings. That allows the tuner to adapt with whatever existing underlying hardware.

\section{Experimentation}
\label{expr}

We constructed a testbed for our if-conversion using \textit{NEAT} to customize the optimization. The operating system is a server version of Ubuntu 14.04 LTS running on top of a multicore system of Intel(R) Xeon CPU E5-2637 v3, which has a speed of 3.5 GHz per core. As for the memory modules, it has a capacity of 64 GB and access rate of 2133 MHz. 

For accurate measurement of time we used Intel's Time Stamp Counter~\cite{rdtsc}
to count processing cycles consumed by the test programs. The implementation of our study methodology was based on the LLVM compilation framework version no. 3.6.1 with our extension that is applicable to any programming language supported by LLVM, and requires no change/addition to the original code of the program. 

We set \textit{NEAT} to form 30 \textit{NNs} per generation and to iterate for 50 generations. These parameters were tuned as a trade-off between the training time -which is proportional to the size of the experiment- and the performance improvement achieved.
As for the test subject, we considered the integer benchmarks from  SPEC-CPU2006 v1.1 \cite{spec} executed with \textit{reference} workload.

\begin{figure*}[!t]
  \centering
  \includegraphics[width=\textwidth]{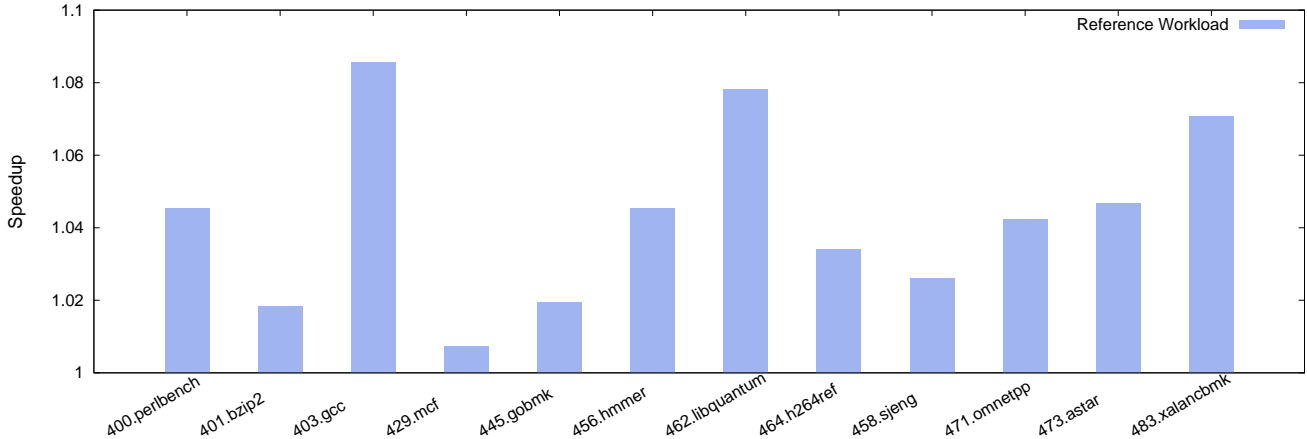}
  \caption{Speedup of benchmarks}%
  \label{fig:speedup}
\end{figure*}

As shown in Figure~\ref{fig:speedup}, our system improves the performance with a percentage ranging from 0.74\% to 8.6\% for all inspected benchmarks over the original if-conversion optimization of LLVM on the same architecture. In Table~\ref{tab:results} we listed the number of branches inspected that can be if-converted while preserving the correctness of the program in every benchmark, in addition to the speedup achieved. As expected, the larger the number of candidate branches, the higher the speedup achieved. \textit{403.gcc} is on the top of list with 4804 branches and 1.086 speedup; on the contrary (\textit{429.mcf}) comes at the end with 12 branches and 1.007 speedup. Moreover, this proportional behavior applies for all the benchmarks except \textit{445.gobmk} and \textit{458.sjeng} which give lower speedups than expected. This is an indication that most of the candidate branches are not part of the frequently executed code which keeps the speedup lower than expected. On the other hand \textit{462.libquantum} gives a much higher speedup compared to the number of candidate branches it contains. This is a clue that these candidate branches are very effective and they have a major influence on the frequently executed code.

\section{Related Work}
\label{related}
The if-conversion related work in the literature is distinguishable by whether if-conversion is performed statically, dynamically, or hybrid. While the previous compiler machine learning-aided techniques target various problems such as the optimal selection of optimizing transformations, their ordering and the best values for the transformation parameters.

Static if-conversion depends principally on information such as the misprediction penalty and the number of cycles within the \textit{if} body; that can be collected in an offline analysis  (profiling) before runtime. 
There are many techniques in the literature that adopt static if-conversion which are described below. 
 \begin{table*}[!ht]
\centering
\begin{tabular}{@{}l c c c @{}}
\toprule
Benchmark&Description~\cite{henning2006spec}&No. Branches& Speedup\\

\midrule
400.perlbench&A cut-down version of Perl
v5.8.7&717&1.045\\
401.bzip2&Compression software based on Julian Seward's bzip2
version 1.0.3&41&1.018\\
403.gcc&C Language optimizing compiler&4804&1.086\\
429.mcf&Combinatorial optimization - Singledepot
vehicle scheduling&12&1.007\\
445.gobmk&Artificial Intelligence - Go game playing&1000&1.019\\
456.hmmer& Search a gene sequence database - Profile HMMs&166&1.045\\
458.sjeng&Artificial Intelligence (game tree search
- pattern recognition)&172&1.026\\
462.libquantum& Physics - Quantum Computing Simulation&25&1.078\\
464.h264ref&Video compression&652&1.034\\
471.omnetpp&Discrete Event Simulation of a large ethernet network&102&1.042\\
473.astar&Computer games - Artificial Intelligence -
Path finding&57&1.047\\
483.xalancbmk& Processor for transforming XML
documents into HTML, text, or other XML document types&934&1.071\\
\bottomrule
\end{tabular}
 \caption{Candidate Branches and Speedup.}%
  \label{tab:results}
\end{table*} 
A compilation framework that delays the if-conversion to schedule time is designed to allow the compiler to minimize runtime by balancing the control-flow and predicated branches~\cite{645801}. 
The authors in \cite{Mantripragada:2000} present an algorithm to perform if-conversion selectively on out-of-order processors that support dynamic speculation and guarded execution. They identified three criteria to measure the profitability of the conversion namely based on size, predictability and profile. The effect of their technique on the net cycles, mispredictions and mis-speculation is exhibited in their paper. 

Another algorithm is reported in~\cite{Snavely02} for the Itanium architecture; it initially operates on unpredicated code, and the if-conversion optimization is performed late in the compilation process. That generates faster (less runtime) and denser (fewer instructions) code. 

Further, an algorithm that minimizes the number of predicates assigned to basic blocks, which are assigned as early as possible using dominance relations to relax dependence constraints, is shown in~\cite{63091}.
Moreover, in~\cite{765952} the program control-flow is represented as a graph (called program decision logic network), then it is modeled in a Boolean equation, which is then minimized and used to regenerate predicated code.

An algorithm that uses dynamic programming to generate code for different target architectures that support predicated execution is discussed in~\cite{hohenauer2008retargetable}.
Another approach handcrafts well-known algorithms (i.e. sorting and searching) into a constant number of branch-free loops, such that a branch predictor can achieve O(1) mispredictions~\cite{aj12}.

Other methods use dynamic if-conversion in which a profiling process during runtime is used to capture some characteristics (e.g. misprediction rate) to be used in optimization.
In ~\cite{hazelwood2000lightweight}, runtime information is used to construct a dynamic optimizer that complements the static one in a previously presented algorithm. It can convert branches, or reverse their conversion, targeting the higher performance based on profiling the program to discover the highly mispredicted branches. Although this algorithm chooses the conversions that improve performance, it does not consider any correlation between the different \textit{if}s which are most probably related to each other especially within the same function.

Also, the authors in~\cite{Jordan13} present two algorithms for if-conversion: one of them targets the intermediate representation (IR) level and the other targets the machine code level. Two heuristics are used to calculate the profitability of the conversion. The
execution time (number
of cycles in the basic block instructions) based on the sizes of basic blocks
and basic blocks execution frequencies.

Also, a hardware that uses runtime information to choose to convert only the hard-to-predict branches is presented in~\cite{kim2006wish}. The presented system provides two versions of the code: one can be predicately executed and the other using a branch predictor.

Finally, there are methods that combine both static and dynamic methods: 
A tree-based model to make predictions using predication and vectorization
techniques are presented in~\cite{DBLP:journals/corr/abs-1212-2287}; it introduces runtime performance ranking assuming that a trained model 
already exists. Meanwhile, data is laid out in memory in an architecture-conscious method. Random features IDs, thresholds, and regression values are used to generate features values in a features vector. 

Another contribution is proposed in~\cite{903263}, where a simple neural network (perceptron) hardware implementation is provided to improve branch prediction. Another method ~\cite{Calder:1997:ESB:239912.239923} uses program-based static branch prediction based on neural networks and decision trees to map
static features associated with each branch to a prediction.

Moreover, an attempt to construct an online ensemble learning framework consisting of small trees to solve the problem of hardware conditional branch prediction is discussed in~\cite{fern03}.
The learning based techniques (the third category) may be the only one that considered the effect of if-conversion on the whole program, as they train their systems over iterations to converge to the least possible runtime. Others make a separate decision for individual \textit{if}s regardless of the correlation with other ones. But, the problem is how fast they can evolve the selection space towards the best performing and that is what we handle in our system.

As an example for research that considered using machine learning algorithms for efficient compiler optimization,  ordering or transformation parameters tuning, we demonstrate some of them below.
A compiler based approach for mapping parallelism to multicore processors is proposed in~\cite{wang2009mapping}. This technique applies an off-line trained \textit{NNs} to develop a number of threads and scheduling predictors for parallel programs. 
An attempt of speeding up iterative compilation -through building models of the program features using machine learning techniques- was presented in~\cite{agakov2006using}.
Authors in ~\cite{monsifrot2002machine} introduced a decision-tree based technique that generates compiler heuristics for the target processor considering the loop unrolling optimization as a source of performance features.
~\cite{yotov2003comparison}.
A per-method logistic regression technique is used to select proper optimizations for each method in the program depending on its features in~\cite{cavazos2006method}.

\section{Conclusion}
\label{conclusion}
If-conversion is a compiler optimization that tackles code performance. It seeks to increase the instruction fetch bandwidth and to decrease branch mispredictions. That can be done by substituting the branches by branch free pieces of code.

This paper investigates the efficacy of a novel technique that uses a machine learning approach that  evolves NNs to solve problems with high complexity, namely \textit{NEAT}; the technique customizes if-conversion in one of the recently developed and rapidly grown open source compilers (LLVM). We examine the adequacy of our system for The INT suite of kernels from the SPEC-CPU2006 v1.1 benchmarks.
 The experiments are performed by extending the LLVM compilation framework to provide for such analysis. The results show notable discrepancy between optimized code using LLVM built-in technique and the best configuration achieved by our system, more than 8.6\% in \textit{462.libquantum}. Moreover, our system does not require neither changes nor additions to the original code of the program. 

Future work would consider dynamic compilation analysis taking into consideration behavior change between calls. On the other hand, we are going to expand our research to include architectures that support predicated execution such as the ARM processors.

\section{Acknowledgement}
We are grateful to Prof. Hironori Kasahara and Prof. Kazunori Ueda, Department of Computer Science and Engineering, Waseda University, Tokyo, Japan, for their contributions, discussions and advice.

\bibliographystyle{abbrvnat}
\bibliography{document}

\end{document}